\documentclass[prd,superscriptaddress,showpacs,amsmath,%
preprintnumbers,showkeys]{revtex4-1}
\usepackage{graphicx}
\usepackage{epsf}
\usepackage{wrapfig}
\usepackage{epsfig}
\usepackage{epstopdf}
\newcommand{\beq}{\begin{equation}}
\newcommand{\eeq}{\end{equation}}
\newcommand\ba{\begin{eqnarray}}
\newcommand\be{\begin{equation}}
\newcommand\ee{\end{equation}}

\newcommand\ea{\end{eqnarray}}
\begin{document}
\title{Photon parton distributions in nuclei and the EMC effect}
\
\author{L~Frankfurt}
\address{School of Physics and Astronomy,
Tel Aviv University,
         69978 Tel Aviv, Israel}

\author{M~Strikman}
\address{Department of Physics,
 Pennsylvania
State University, University Park, PA 16802, USA
}
\date{\today}
\begin{abstract}
Photons as well as quarks and gluons are constituents of the infinite momentum frame 
(IMF) wave function of an energetic  particle. They  are mostly equivalent photons whose 
amplitude follows from   the Lorentz transformation of  the particle rest frame Coulomb 
field into  the  IMF  and from the conservation of the  electromagnetic current.  We evaluate in a model independent way the dominant  photon contribution   $\propto \alpha_{em}(Z^2/A^{4/3})\ln(1/R_{A}m_{N}x)$
to the nuclear structure functions as well as the term $\propto \alpha_{em}Z/A$. 
In addition we show that the definition of $x$ consistent with 
the exact kinematics of $eA$ scattering (with exact sum rules)
works in the same direction as the nucleus field of equivalent photons. 
Combined, these effects account for the bulk of the  EMC effect for $x\le 0.5$ where Fermi motion effects are small. In particular for these x  the hadronic mechanism contribution to the EMC effect does not exceed 
$\sim 3\%$ for all nuclei. 
Also the A-dependence 
of the hadronic mechanism of the   EMC effect for $x >  0.5$ is significantly modified.
 \end{abstract}
\maketitle

\section{Introduction}

Atomic nuclei carry electric charge. Therefore the Coulomb field of a nucleus is a
fundamental property of the nucleus in its rest frame. Under the  Lorentz transformation 
to the frame where nucleus  has a large momentum, the   rest frame nucleus Coulomb field  is 
transformed into the field of equivalent photons. This phenomenon is well known as 
Fermi - Weizsacker - Williams  approximation for  the wave function of a rapid projectile with 
nonzero electric charge \cite{WW}. For high energy processes where the 
photon is off mass, 
shell this result has been generalized by V.~Gribov   \cite{GribovComplex}. 
Application of these methods allows  the evaluation of   the role of photon degrees of freedom 
in the partonic nucleus structure. This effect has practical
 implications for A-dependence of the EMC 
effect for $A\ge 50$ nuclei, for  the hard processes with the violation of isotopic 
symmetry in the DIS off protons and neutrons and nuclei. It also gives  a non-negligible contribution into 
extraction of the Weinberg angle from the (anti)neutrino scattering off iron which was performed
by the NuTeV experiment at Fermilab, see \cite{nutev} for a review and references (implications of our finding for this effect 
will be considered elsewhere). Another effect that modifies the A-dependence of the EMC effect 
is the difference between the  conventional definition of $x=AQ^2/2M_Aq_0$ for the scattering off 
nuclei which is consistent with the energy momentum conservation sum rule and the one used in experimental papers - $x=Q^2/2m_p q_0$. Combined these two effects result in a strong reduction 
of the hadronic contribution to the EMC effect for  $x \le 0.5$ where modification of $F_{2A}(x,Q^2)$ due to the  nucleon  Fermi 
motion is small.

The photon component of the parton distribution functions  of nucleons has been  considered for a long time, see for example \cite{Dokshitzer:1978hw,Gluck:1994vy,Gluck:2002fi}. However the emphasis was on the deep inelastic contribution due to the $Q^2$ evolution of the quark distributions.  In particular the current MRST analysis \cite{Martin:2007bv} includes only such contributions neglecting the 
contribution of equivalent photons to the photon structure function of  nucleons,
 that is, contribution of  a photon plus the  nucleus ground 
 state.

It has been observed experimentally that nuclear structure functions differ from the sum of 
structure functions of nucleons. At small $x$ this is the celebrated nuclear shadowing effect 
predicted within the preÐquantum chromodynamics (QCD)  framework in \cite{Gribov}   and evaluated in the leading twist 
approximation in  \cite{FS88,FS} and 
observed in a number of experiments, see review  in \cite{data}.  The observed difference between 
nucleus structure functions at medium $x$ and the sum of that for free nucleons is known 
as the EMC effect, for a review  see  \cite{data}. The presence of the EMC effect requires the presence of non-nucleonic degrees of freedom
in nuclei.    A 
natural mechanism is a deformation of the  distribution of color within the wave functions of the 
bound nucleons, for a  recent  discussion and references see \cite{CFKS}.  One of the non-nucleonic 
effects is the presence of  photon degrees of freedom in the light- cone wave function of a 
nucleus. Note that structure  functions of a hadron target are calculable in terms of its light-cone 
wave functions \cite{Feynman}  and their QCD evolution.

The paper is organized as follows. In Section 2 we review  the general framework for treating the photon 
field as a a constituent of the parton wave function of the nucleus. In Section 3 we calculate the 
contribution of the equivalent photons due to the  coherent nucleus final state as well as corrections 
due to the incoherent final states.  Numerical results are presented in Section 4 where we also 
consider another effect contributing to the EMC ratio - proper definition of Bjorken $x$ for the 
scattering of nuclei that  is consistent with the momentum sum rule. We 
demonstrate 
that  when combined, these two effects account for  the bulk of the l EMC effect for $x\le 0.5$. We 
also include in the lowest order in $k^2/m_N^2$ the effect of nucleon Fermi motion which becomes increasingly important with an increase of x at $x > 0.5$. 
We conclude that the major  "hadronic contribution " to the EMC effect  is concentrated at $x > 0.5$. In Section 5 we  outline implications of our analysis for the global fits of nuclear parton distribution functions (pdf's). Our conclusions are presented in Section 6.

\section{Photon distribution in  heavy nuclei.}

A nucleus is characterized by quark, gluon, photon  distributions within a nucleus.   To suppress 
particles production by hard probe from vacuum   the gauge condition $A^{+}=0$ is chosen,  where $A_{\mu}$ is the operator of photon field.
In this gauge  photon distribution has  a  form familiar from the parton model; compare \, Fig.~1.
 The photon parton distribution can be written as the matrix element of the product of the operators 
cf.  \cite{Sterman}:
\begin{equation}
P_{A}(x_A,Q^2)=(2\pi x_Ap^{+})^{-1} \int_{-\infty}^{\infty}dy^{-} \exp{(-ix_Ap^{+}y^{-})}
1/2 \sum_{\mu}<A\left|\left[F^{+}_{\mu}(0,y^{-},0),F^{\mu ,+}(0)\right]\right|A>_{A^{+}=0}.
\label{parton1}
\end{equation}
Here  $F_{\mu,+}$ is the operator of the strengths  of the photon field with transverse 
component $\mu$,    and $x_A=Q^2/2(p_Aq)$ is the Bjorken $x$ for the nuclear 
target which is 
the fraction of the nucleus momentum carried by a parton (we will later rescale it by a factor of A to match it more closely to the case of the nucleon target).

\begin{figure}[t]  
   \centering
   \includegraphics[width=0.7\textwidth]{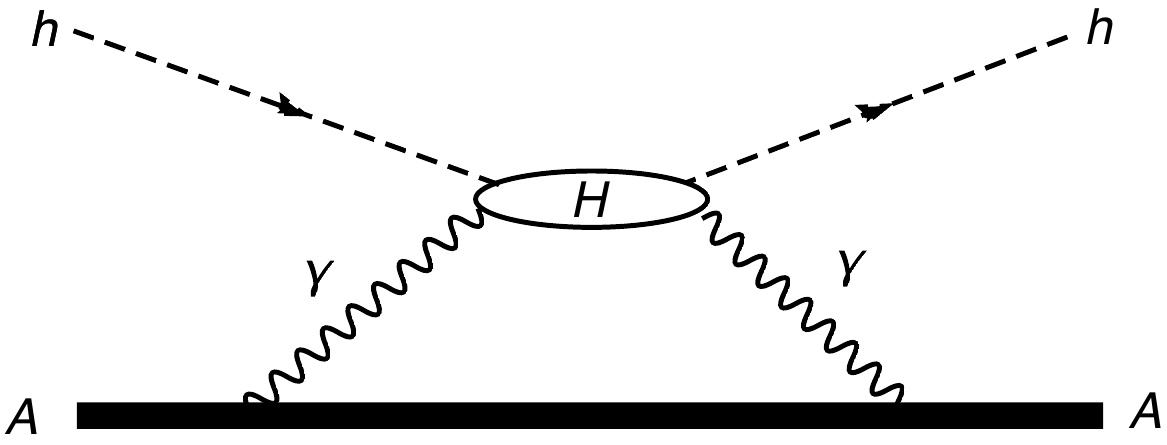} 
   \caption{Diagram for the interaction of photon of the nucleus with a hard probe $h$}
    \label{plot}
 \end{figure}

A large group of hard processes  will probe high energy processes off
the nucleus Coulomb field in the  ultraperipheral processes at the 
LHC\cite{Baltz:2007kq}.  One example of ultraperipheral processes is high 
energy photon scattering off nucleus(nucleon) Coulomb field  with diffractive 
production of massive lepton pairs $\gamma+A\to L^{+}+L^{-}+A$.
Another  process  involving photon distribution is  the  exclusive 
meson photoproduction off the Coulomb field of the nucleus (the Primakoff effect).

To simplify the calculations it is convenient to represent $P_A$  as the sum of two contributions  $P_A=P^{inel}+P_A^{coherent}$.  The first term includes  nucleus  excitations  - we will refer to it as the  inelastic term. 
The second  term is  the contribution of equivalent photons.
We will refer to it as the coherent  term.

Let us start with a brief discussion of the inelastic contribution.  Above we 
defined  it  exactly the same way as other parton densities,  cf. Eq.\ref{parton1}.
This definition allows to calculate 
$P_A$ by evaluating  corresponding Feynman diagrams, cf. Fig.~1.
\begin{equation}
x_AP^{inel}_A(x_A,Q^2)={\alpha_{em}\over \pi} \int d^2 k_t  \int _{\nu_{min}}^{\nu} {d\nu'\over \nu'}
  {k_t^2 \over (k_t^2+Q^2\nu'/\nu)^2} F_{2A}(\nu',k_t^2).
\label{photon}
  \end{equation}
Here $\nu=2(pq)=Q^2/x_A$,  $k_t$ is the transverse momentum of the photon, 
and $F_{2A}(\nu,Q^2)$ is the nucleus structure function which does not include 
the photon field.  In the above formulas  we neglected  the small contribution of  
the  longitudinally polarized  photons ($F_L^A$).The calculation of the photon 
structure function  accounting for the nucleus excitations is essentially the  same 
as for the QCD evolution of the gluon distribution, cf.\cite{Sterman}.  The  
analogous expression for the photon distribution in protons and neutrons is given 
in \cite{Gluck:2002fi}.

Account of  the photon degrees of freedom requires modification of the  QCD 
evolution equation by including $x_AP_A$ in addition to $x_AG_A, x_AV_i,x_AS_i$.  
The presence of the photon component 
in the nuclear light-cone wave  function  leads to the modification of
the  momentum sum rule as follows
 \begin{equation}
 \int_0^1 \left[x_{A}V_{A}(x_{A}, Q^2) +x_{A}S_{A}(x_{A}, Q^2) +x_{A}G_{A}(x_{A},Q^2)+ x_{A}P_{A}(x_{A}, Q^2)\right]dx_{A}=1.
 \end{equation}
To remove the  kinematic effects 
it is convenient to redefine the  variables by introducing  
\begin{equation}
x=Ax_{A}.
\label{xdef}
\end{equation}
leading to 
 \begin{equation}
 \int_0^A \left[(1/A)(xV_{A}(x, Q^2) +xS_{A}(x, Q^2) +xG_{A}(x,Q^2))+xP_{A}(x, Q^2)\right]dx=1.
 \label{sumrule}
 \end{equation}
We will show that in the case of heavy nuclei  the photon distribution  in a nucleus
cannot be neglected  in the evaluation of the EMC effect.

The model for  the evaluation of impact of  the internucleon electromagnetic 
interactions (which contribute to the nuclear  binding energy term $\propto Z^2/R_A$)
 into the momentum 
sum rule  for the heavy nucleus target  has been suggested  in   
Ref. \cite{Birbrair:1989zf}.   The model suggested a  relation between the  nucleus Coulomb 
binding energy and the  momentum sum rule  through  the application of the virial  theorem in  
the nucleus  rest frame.  So far there exists no model independent derivation of the 
connection between rest frame  Coulomb energy and the momentum sum rule.  
Their model estimates differ from the well established  equivalent photon approximation employed 
 in this paper. Our numerical results are also different.
  
\section{The contribution of   equivalent photons}

Here we will calculate the coherent contribution to  the parton nucleus distribution  
that  dominates 
the photon distribution in a nucleus in the leading order in $Z$. This contribution to 
$P_A(x,Q^2)$ arises from the interaction of a hard probe with a photon coherently emitted by   the target nucleus.   The coherent contribution to the photon structure function  is unambiguously calculable  in terms of the electromagnetic form factors 
of  the nucleus target.  In the calculation we neglect by the small contribution of the magnetic form factor of the nucleons which is concentrated at large $k^2_t$.

We calculate the field of equivalent photons due to  the Lorentz transformation of 
the familiar nucleus rest frame Coulomb  field. To achieve fast track in the 
calculation of the light-cone nucleus wave function we explore below the 
relationship between $x$  and the  nucleus momentum  in the intermediate  state. 
Calculations are simplified in our case since the  nucleus is heavy so the static approximation 
should be sufficiently accurate. In the static approximation  zero component of photon 
momentum in the nucleus rest frame is negligible:  $k_0=k^2/2m_A$.  So
\begin{equation} 
 x=A(k_0-k_3)/M_A\approx -k_3/m_N.
 \end{equation}

The second simplification arises from the observation that the four-vector  $k$ can be decomposed over  directions defined by external momenta: 
$k_{\mu}=ap_{\mu}+bq_{\mu}+k_t$.    Here $p$ is the four-momentum of the target 
nucleus and $q$ is the four-momentum of  the virtual photon (external 
hard probe) and $(p k_t)=(q k_t)=0$.  In the essential region:   
$p_{\mu}A_{\mu, \lambda}\approx (k_{t \mu}/a) A_{\mu,\lambda}$the essential region, p?A?,? Å (kt?/a)A?,? \cite{GribovComplex}. Account of this property leads to the generalization 
of the Fermi - Weizsacker -Williams expression 
for the spectrum of the equivalent photons:
\begin{equation}
xP^{coherent}_{A}(x,Q^2)={\alpha_{em}\over \pi } {Z^2\over A}
\int  k_t^2 d^2 k_t \frac{F_A^2(k_t^2+x^2 m_N^2)}{(k_t^2+x^2 m_N^2)^2},
\label{ww1}
\end{equation}
For the characteristic  values of  $k_t^2$ in this integral which are determined  by the nuclear 
form factor it is legitimate 
to neglect   the  proton form factor. The finite size of the  nucleus can be 
accounted for by the nuclear electric form factor  $F_A(t)$, and in our  
estimates we choose $F_A$ in the exponential form, that is, 
\begin{equation}
 F_{A}(k_t^2+x^2 m_N^2)=\exp(-R_A^2 (k_t^2+x^2m_N^2)/6).
 \end{equation}
\noindent
Here $R_A$ is the RMS nuclear radius and $m_N$ is the  nucleon mass.
 
Such a form  allows one  to perform the  integration over  the transverse momenta of 
the photons and to calculate the leading logarithmic term in the essential region of small $x$.
We obtain:
 \begin{equation}
xP_A^{coherent}(x)=
  \frac{2 \alpha_{em}Z^2}{A} \ln \left(\frac{\sqrt{3}}{R_Am_N x}\right) 
  \exp \left(-{R_A^2m_N^2x^2\over 3}\right).
\label{P}
\end{equation}

Note,  for  completeness, that the  contribution of equivalent photons would dominate  
the nucleus structure function in the limit of fixed $x$ and  $Q^2\to \infty$.  This is because 
the  coherent contribution is practically $Q^2$ independent 
since form factor of the photon is a slow function of $Q^2$  due to smallness of $\alpha_{em}$.  
In the case of  the lightest nuclei the $P_A$ is too  small due to the  smallness of the electromagnetic 
coupling  constant $\alpha_{em}$ to lead to noticeable effects.  In the case of medium and 
especially heavy nuclei quantity $Z\alpha_{em}$ is not small, so that the photon  becomes 
an important constituent of the light cone  wave function of the nucleus.

It is useful to compare different electromagnetic contributions.  The most important
one is the contribution of equivalent photons in which  the fields of individual protons add 
coherently.   This coherence leads to a larger momentum fraction carried by the photon 
field in nuclei as compared to that carried by  individual free protons.
Approximate Eq.~\ref{P} can be used to evaluate   the coherent contribution of the photons to 
the momentum sum rule:
 \begin{equation}
 \lambda_{\gamma}^{coherent}={2\alpha_{em}}  {Z^2\over A}  \int\limits_0^1 dx \ln (\sqrt{3}/x m_NR_A)
\exp(-R_A^2m_N^2 x^2/3)\approx
 {\alpha_{em}}  {Z^2\over A}
 {\sqrt{3}\over m_NR_A}.
 \label{ww5}
 \end{equation}
 \noindent
 Therefore the incoherent contribution  to the  momentum sum rule $\propto \alpha_{em} Z/A$
 can  be safely neglected except for the light nuclei. 
 
 A direct integration of Eq. \ref{ww1} leads to 
 \begin{equation}
 \lambda_{\gamma}^{coherent}=\int_0^1dx xP_{\gamma}(x,Q^2)={\alpha_{em}}  {Z^2\over A}
 {1.759 \over m_NR_A}.
 \label{ww6}
 \end{equation}
  which differs from an approximate result of Eq.\ref{ww5} by a factor 1.0156.

  \begin{figure}[t]  
   \centering
   \includegraphics[width=0.8\textwidth]{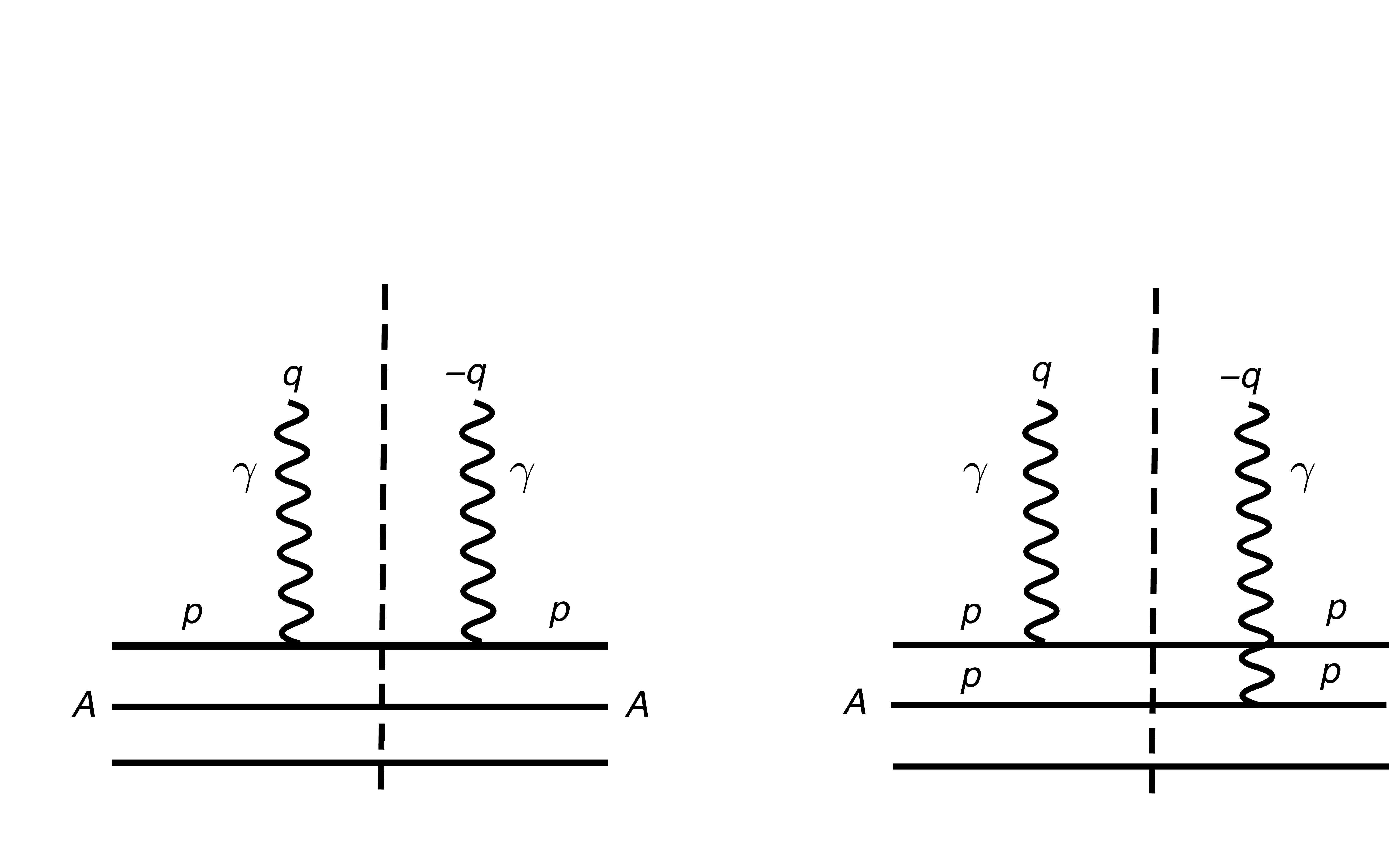} 
   \caption{Diagrams for the photon component of the nucleus including
   coherent contribution of the whole nucleus and incoherent contributions of the photon fields of individual protons.}
    \label{double}
 \end{figure}

  In the case of light nuclei we need also to take into account 
 contribution of the incoherent break up of the nucleus. 
 Sum of the two effects can be calculated in the closure approximation 
 where cross section is described by the sum of two diagrams presented in Fig.\ref{double}
 (cf. \cite{Frankfurt:2000jm}) and results in  replacing 
 \begin{equation}
 Z^2F_A^2(t) \longrightarrow  ZF_N^2(t) + Z(Z-1)F_A^2(t),\label{replace}
 \end{equation}
 in Eq.\ref{ww1} (the first and second terms correspond to Fig. 2a, and Fig.2b). Here $F_N(t)$ is 
 the proton electric  (Dirac) form factor and $F_A(t)$ is the observed electric form factor of the 
 nucleus which is equal to the product of the nucleus body form factor and $F_N(t)$. The  first 
 term in Eq.\ref{replace} is due to the Coulomb field of the individual protons and it is included 
 in the structure functions of the proton.  The contribution of magnetic form factors of neutron and 
 proton is negligible because it is concentrated at $-t$ larger than the scale of nuclear phenomena.  
 Hence to calculate the additional fraction of the momentum carried by photons in the nuclei we 
 simply need to change $Z^2$ to $Z(Z-1)$ in Eqs.\ref{ww1},\ref{P} - 
 \ref{ww6}.  In particular we obtain
 \begin{equation}
 \lambda_{\gamma}=\int_0^1\, dx xP_{\gamma}(x,Q^2)={\alpha_{em}}  {Z(Z-1)\over A}
 {1.759 \over m_NR_A}.
 \label{ww7}
 \end{equation}

  Taking $R_A$ from the compilation of ref.\cite{Angeli} we find
  \begin{equation}
\lambda_{\gamma}(^4He) = .08\%;\, \lambda_{\gamma}(^{12}C) = .27\% ;\, \lambda_{\gamma}(^{27}Al) = .51\% ;
 \lambda_{\gamma}(^{56}Fe) =.84\%;
\, 
\lambda_{\gamma}(^{197}Au) =1.56\%.
\end{equation}

To evaluate the impact of the presence of the photon  component for the nuclear  structure 
functions we can use   a  reasonable starting approximation:
$F_{2A}=ZF_{2p}+(A-Z)F_{2n}$. The small x nuclear shadowing effects modify this 
approximation, however they are negligible for $x\ge 0.2$ range we are interested in. Since  
deviations from the additivity are small the effect of the presence of the photons and other 
"hadronic" effects can be treated as contributing additively to
the deviation of the EMC ratio (Eq.\ref{emcr}) from one.

\section{Implications for the EMC effect}

\subsection{Contribution of equivalent photons}
The light-cone momentum carried by the photons is compensated by
the loss of the momentum by the nucleons.  Since the  Coulomb field generated by the protons is soft, it is natural to assume that 
reduction of the light cone fraction experienced by protons is shared with neutrons due to the internucleon interactions that typically change light cone fractions by a larger amount than the overall change of the light cone fraction due to the photon field. Hence we will assume that the shift is approximately equal for protons and neutrons (an assumption that all shift is due to protons leaving the neutron distribution mostly unchanged leads to a slightly larger EMC effect).

Neglecting other sources of the EMC effect, it is easy to demonstrate that the
suppression  effect at not too large $x$ can be expressed through  the value
of the fraction carried by photons as  follows 
\begin{equation}
R_A(x,Q^2) =  {ZF_{2p}(x/(1-\lambda_\gamma),Q^2) +NF_{2n}(x/(1-\lambda_\gamma,Q^2)  \over
ZF_{2p}(x,Q^2) + NF_{2n}(x,Q^2) }.
\label{emcr}
\end{equation}
Experimentally the ratio is defined relative to $F_{2^2H}(x,Q^2)$:
\begin{equation}
R_A(x,Q^2) =  {F_{2A}(x,Q^2) \over
F_{2^2H}(x,Q^2) }.
\label{emcr1}
\end{equation}
where factor $A$ in $F_{2A}(x,Q^2)$ is explicitly taken out. 
Since  the Coulomb effect is negligible in the deuteron case we can ignore this difference of the definitions. Also, the SLAC experiment \cite{Gomez:1993ri} introduced a correction for an unequal number of protons and neutrons. Hence in spite of he uncertainties in this procedure for the sake of comparison with the data it is reasonable to treat nuclei as isoscalar  targets.

Using the Taylor series expansion, we obtain
\begin{equation}
R_A(x,Q^2) -1 = - {\lambda_{\gamma}xF_{N}^{\prime}(x,Q^2)
 \over F_{N}(x,Q^2)
 }.
\end{equation}
Parameterizing
$F_{2p}(x)+F_{2n}(x) \propto (1-x)^n, n\sim 3$ we obtain:
\begin{equation}
R_A^{Coulomb}(x,Q^2)=1 - \lambda_{\gamma} {nx\over (1-x)}. \label{enha}
\end{equation}
The results of calculations  Eq.\ref{enha} are  presented 
in the first column of Table ~1. 

\subsection{Account of kinematics of DIS}

Since the EMC effect is small it is legitimate to consider it as the sum of 
different effects that  can be investigated separately. One
small but noticeable effect is 
due to a proper definition of the Bjorken x. In the 
impulse approximation in case of restriction by the nucleon degrees of freedom in the nucleus wave function one needs to perform comparison of the cross sections for the same $x$ defined 
in Eq.~\ref{xdef}, since this $x$ enters in the  convolution formulas for the cross section
for the nucleon light-cone fraction $\alpha$ scaled to vary between 0 and A:
\begin{equation}
F_{2A}(x,Q^2)= \int {d\alpha\over \alpha} d^2p_t F_{2N}({x\over \alpha},Q^2) \rho_A^N(\alpha,p_t),
\end{equation} 
where $\rho_A^N(\alpha,p_t)$ is the light-cone density matrix. $\rho_A^N$ satisfies momentum and baryon charge rules which follows from
the conservation of energy-momentum and baryon charge, see discussion in \cite{FS88}.   

At  the same time the experimental data 
are presented for the same $x_p=Q^2/2m_pq_0$.
The ratio 
\begin{equation}
x/x_p=Am_p/m_A= (1+ (\epsilon_A - (m_n-m_p) N/A)/m_p),
\end{equation}
where $\epsilon_A $ is the energy binding per nucleon.
Using as before the Taylor series expansion we find for the resulting  correction for the ratio of nuclear and deuteron cross sections:
\begin{equation}
R_A^{\epsilon} = F_{2A}(x_p)/ F_{2^2H}(x_p) = 1 - {nx\over 1-x} (\epsilon_A- \epsilon_{^2H}  - (m_n-m_p) (N-Z)/A)/m_p,
\end{equation}
We use the observation that in the kinematical region of $x < 0.55$ Fermi
motion effects are a negligible correction and therefore 
$F_{2A}(x,Q^2)=F_{2N}(x,Q^2)$.

The two effects add as both are small corrections. 
They  lead 
to the overall contribution to the EMC ratio  
which is not due to hadronic non-nucleonic degrees of freedom in nuclei 
\begin{equation}
R_A(x_p,Q^2) = F_{2A}(x_p)/ F_{2^2H}(x_p)= R_A^{\epsilon}(x_p)\cdot  R_A^{Coulomb}(x_p). 
\label{sum}
\end{equation}

\subsection{Comparison with data}

\begin{figure}[b]  
   \includegraphics[width=0.6\textwidth]{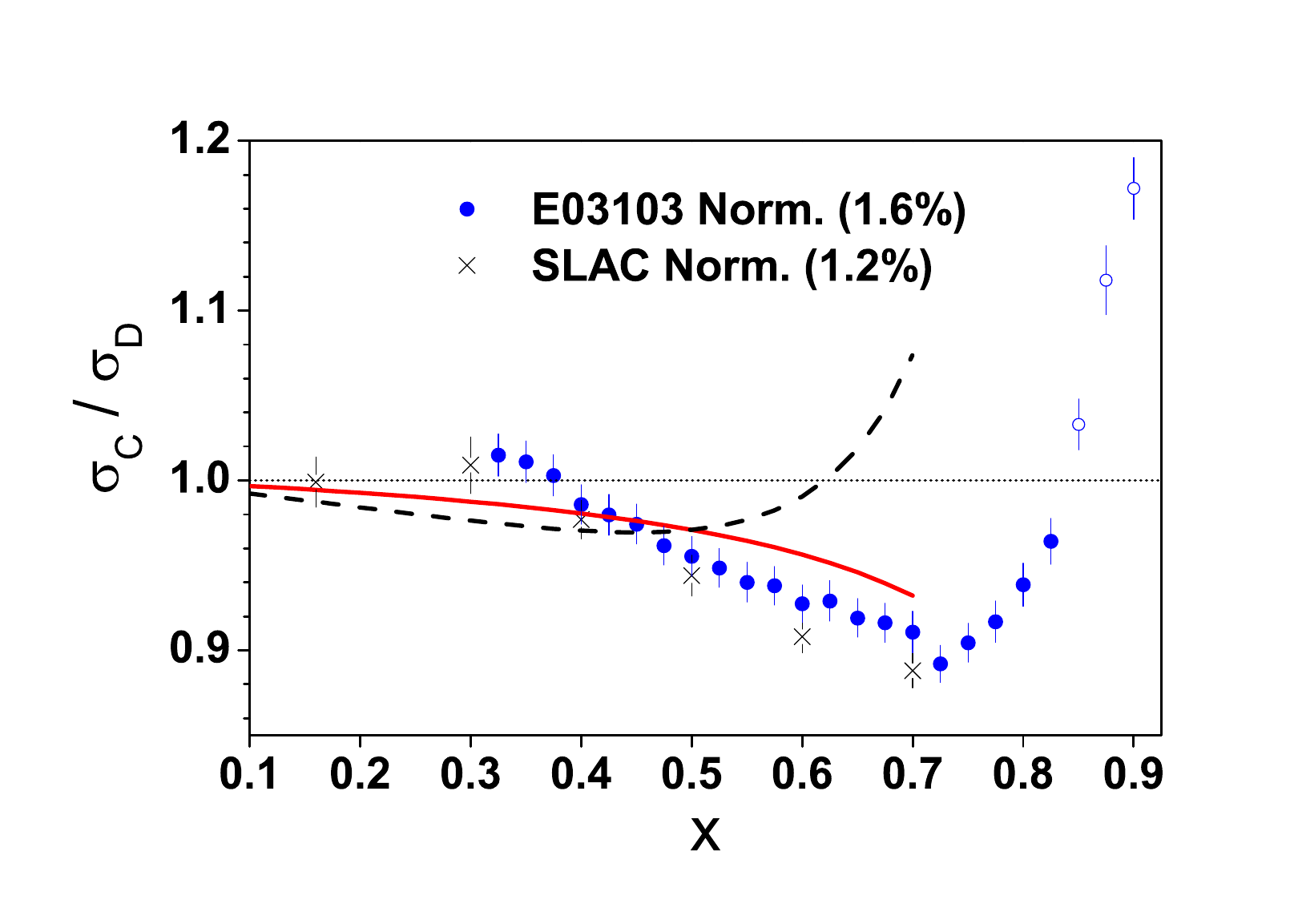} 
   \includegraphics[width=0.6\textwidth]{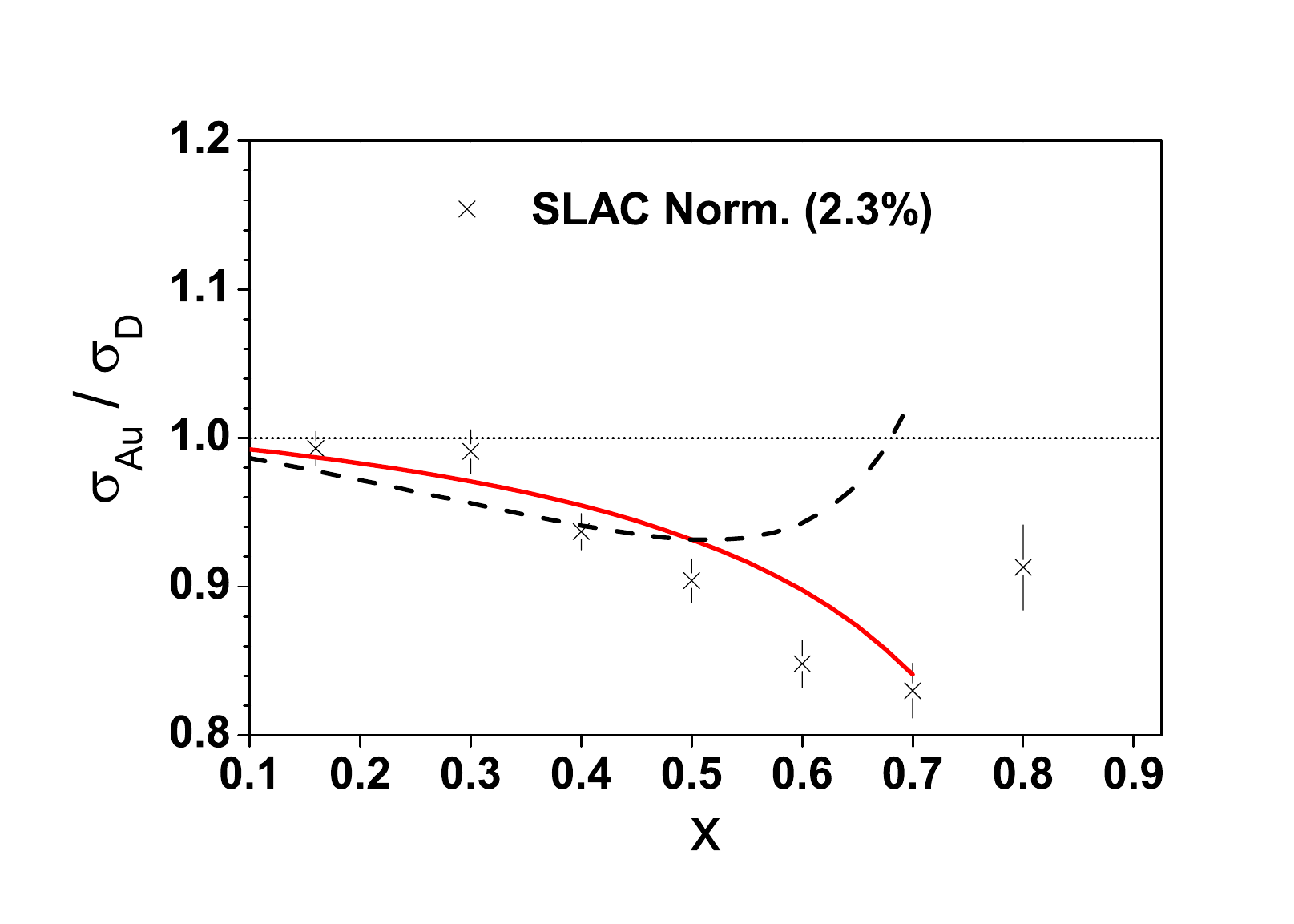}
   \caption{The solid curve is the  contribution to the EMC ration of 
   the nucleus field of  equivalent photons and effect of proper definition of x calculated using Eq.\ref{sum} which is applicable for $x\le 0.7$ only.
   The dashed curves include also the  effect of the Fermi motion estimated using Eq. \ref{Fermi}.
   The data are from  \cite{Gomez:1993ri,Seely:2009gt}.   Open circles correspond to $W< $2GeV.}
    \label{data}
 \end{figure}

The results of the calculation are presented in the first two  columns of Table~1.
The third  and forth columns present the experimental results   
 of \cite{Gomez:1993ri} and  \cite{Seely:2009gt}.  Results for the carbon and gold targets are also presented as solid curves in Fig. \ref{data}.

One can see both from Fig.\ref{data} and the Table 1 that    the discussed effects change strongly the A-dependence and the strength of the 
"hadronic mechanism"  of the EMC effect. In particular, the difference between  the 
equivalent  photon contributions for $^{12}$C and $^4$He is  comparable 
to the observed difference between the EMC effect for these two nuclei while the strength of the "hadron mechanism" is at least a factor of 2$\div $ 3 smaller for $x=0.5$.

\begin{table}[!hc]
   \begin{center}
   \caption{The  contributions to the EMC ratio for x=0.5, x=0.6 not related to hadronic non-nucleonic degrees of freedom: (a) Coulomb contribution  (b) Combined effect of the Coulomb contribution and proper definition of $x$, (c)   the data of  [19]. First error is combined statistical and systematic error, the second is the overall normalization error. (d)  the data of  [20]. First error is  statistical, second one is  systematic error, the third one is the overall normalization error.}
     \begin{tabular}
     {|c||c|c|c|c|}\hline
    &   Eq. (\ref{enha}) &  Eq. (\ref{sum}) & [19] & [20] \\\hline
$^4$He & 0.998 & 0.979 &0.949 $\pm$ 0.016  $\pm$ 2.2\%& 0.9695 $\pm$ 0.0060 $\pm$ 0.0099 $\pm$ 1.5\%\\\hline
$^{12}$C   & 0.992  & 0.971 & 0.944 $\pm$ 0.010  $\pm$ 0.7\% & 0.9553 $\pm$ 0.0050$\pm$ 0.0106 $ \pm$ 1.6\%\\\hline
$^{27}$Al   & 0.985  & 0.962 &0.930 $\pm$ 0.008  $\pm$ 0.7\% &\\\hline
$^{56}$Fe  & 0.975 & 0.950 &0.911 $\pm$ 0.007  $\pm$ 1.0\% &\\\hline
$^{197}$Au   & 0.953 & 0.932 & 0.904 $\pm$ 0.009  $\pm$ 2.3\%& \\\hline
     \end{tabular}
     
     $R_A(x=0.5)$
     
     \vspace{1cm}
         \begin{tabular}
     {|c||c|c|c|c|}\hline
   &   Eq. (\ref{enha}) &  Eq. (\ref{sum}) & [17] & [18] \\\hline
$^4$He & 0.996 & 0.968 & 0.962 $\pm$ 0.016  $\pm$ 2.2\%& 0.9491 $\pm$ 0.0043 $\pm$ 0.0099 $\pm$ 1.5\%\\\hline
$^{12}$C   & 0.988  & 0.956 &0.908 $\pm$ 0.007  $\pm$ 0.7\% & 0.9274 $\pm$ 0.0039 $\pm$ 0.0103 $ \pm$ 1.6\%\\\hline
$^{27}$Al   & 0.977  & 0.943 &0.904 $\pm$ 0.007  $\pm$ 0.7\% & \\\hline
$^{56}$Fe  & 0.962 & 0.927 & 0.874 $\pm$ 0.007  $\pm$ 1.0\% &\\\hline
$^{197}$Au   & 0.930 & 0.898 & 0.848 $\pm$ 0.008  $\pm$ 2.3\%&  \\\hline
\end{tabular}

$R_A(x=0.6)$

 \end{center}
\label{Tab1}
\end{table}

The comparison of the calculation with the data at $x=0.6$ 
shows that "the hadronic EMC effect " 
leads to $R_A(x)$ close to one. Note however that data indicate
a significant deviation from a naive expectation of the impulse approximation
where $R_A (x)> 1 $  for $x>0.5$ 
and growing with increase of $x$. 

Indeed, let us consider the effect of the Fermi motion of the nucleons in nuclei. For moderate $x \le 0.7$ one can write the contribution of this effect to the nucleus/ deuteron ratio as \cite{FS88}:
\begin{equation}
R_A(x,Q^2) =1 +  {xF_{2N}^{\prime}(x,Q^2)+(x^2/2)F_{2N}^{\prime\prime}(x,Q^2)\over F_{2N}(x,Q^2)} \, \cdot {2(T_A-T_{^2H})\over 3m_N},
\end{equation} 
where $T_A$ is the average nucleon kinetic energy. Taking as before 
$F_{2N}(x,Q^2) \propto (1-x)^n$ we obtain
\begin{equation}
R_A(x,Q^2) =1 + {n x  (x(n+1) -2) \over (1-x)^2}\cdot 
{(T_A-T_{^2H})\over 3m_N}.
\label{Fermi}
\end{equation} 
Including this effect using estimates of $T_A$ from \cite{CFKS}
as well as  the effects we considered before leads to the dashed curves in Fig.\ref{data}.

One can see from Fig.\ref{data} that deviation of the data from the dashed curves grows with increase of $x$ for $x\ge 0.6$ indicating increasing importance of the "hadronic EMC effect". 
This  is consistent with the expectation that suppression of the EMC ratio due to the effect of the suppression of the point-like configurations in bound nucleons should be maximal for quarks that  carry a fraction of the light - cone momentum of the bound nucleons close to 1 \cite{FS88}.  Since the Fermi motion effect is  $\propto T_A$ for $ 0.5 < x \le 0.7$, the compensating 
 "hadronic EMC effect " should also be approximately proportional to $T_A$.
 For the realistic nuclear wave functions $T_A$ is dominated by the contribution of the short-range $pn$ correlations giving further support to the expectation that modifications of the bound nucleon structure are maximal for the short-range correlations.
 
Note also that many features  of  the EMC effect due to the photon field found in the paper are 
rather similar  to the pattern  of  the pion model of the EMC effect,  see discussion of this model 
in \cite{FS88}. Thus  account of the nucleus photon field puts a stronger limit on the possible  
contribution to the EMC effect  of the  suppression of 
nuclear structure functions due to deformation of the  nucleus pion field.

\section{Implications for global nuclear pdf analyses}

The observed effects obviously impact on the global analyses of the nuclear pdfs.
Here we briefly outline several of the effects.

{\it (a) Correction for the gluon pdfs}

\noindent 
All analyses were imposing the momentum sum rule (\ref{sumrule})
without including  the photon field. This leads to an overestimate of the fraction of the momentum carried by gluons. Since the charged partons  carry approximately 50\% of the momentum of nuclei, the correction for the contribution of the gluons to the momentum sum rule $\int_0^A {1/A}xg_A(x,Q^2) dx$ is $\approx  1- 2 \lambda_\gamma$. For the case of heavy nuclei, like lead,  which are used in
the heavy ion collisions experiments, the correction is $\approx 3.3\%$. Half of  the effect is accounted for by  a rescaling of $x$ by a factor of $(1+\lambda_{\gamma})$. The rescaling  mostly changes the large x gluon distribution. For small x effect is on the scale of 1\%.

{\it (b) Correction for the valence quark pdfs}

\noindent
The current applications of the baryon charge sum rule link the large $x$  suppression of $V_A(x) $ to the enhancement of $F_{2A}$ at $x \sim 0.1$. However the effects we discussed above lead to compensation of   the depletion by enhancement at  much smaller x where $ xV_A(x)$ decreases with 
decrease of $x$. This leaves more room for the effect of the leading twist nuclear shadowing which tends to compensate the leading twist shadowing by a corresponding enhancement at $x \sim 0.1$, see discussion in \cite{FS88}.

{\it (c) Correction for the sea  quark pdfs}

For the case of the antiquark distribution, 
$x\bar q(x) \propto (1-x)^n, n\sim 7$ application of Eq.\ref{sum} leads to a suppression factor for the highest  $x\sim 0.2 (0.25)$ reached in the Drell-Yan experiments with heavy nuclei of about $R_{\bar q}= .96 (.95)$ and to even larger effect for the forthcoming Drell - Yan experiment at FNAL.

Overall, it is clear that a new global analysis of the data including discussed effects  is  necessary.

\section{Conclusions}

We demonstrated that the photon component of the light-cone wave function of the nucleus 
gives a non-negligible contribution to  structure functions of heavy nuclei.  Our 
formula for the  photon distribution is a kind of the LO QCD evolution equation and  therefore it is exact within the LT approximation,  cf. Eq.\ref{photon}.
Account for the nucleus photon field
leads to a significant EMC effect  modifying the A-dependence of the hadronic 
contribution to the EMC effect. Account of photon field
is important for interpretation of both relative strength of the EMC 
effect for $x\ge 0.5 $ - for example $^4$He  vs $^{12}$ C,  as well as for evaluation of the overall A-dependence of the hadronic mechanism of the EMC effect. We also find a significant modification 
of the EMC ratio due to a proper definition of the Bjorken $x$. Combined,
the  two effects account for most  of the EMC effect for $ x \le  0.5 \div 0.6$. 

Subtraction of the contribution of photon component of light-cone nuclear wave function  
slows down the dependence of the EMC effect $\left[R_A(x)-1\right]$ on the atomic number for $x\sim 0.5 \div 0.6$ making it closer to the A dependence of the  the probability of short range nucleon 
correlations in nuclei, and hence  more close to the expectations of color screening model of \cite{FS88} in which radii of  quark, gluon orbits in
short-range correlations increase with increase of the nucleon momenta.

It is important to perform further experimental measurements of the EMC ratio for $x\ge 0.6$ at large 
$Q^2$ which will become feasible at Jlab 12. Especially interesting would be to study the ratio of the 
EMC effect in $^{48}$Ca and $^{40}$Ca since in this case both effects we consider are practically the same.   Moreover such research would  provide a unique window on the structure of neutron rich high density nuclear matter relevant for description of the cores of the neutron stars.

 It would be possible to measure small x structure functions of 
nuclei  and therefore photon  distributions via  Eq. \ref{photon}  
at the LHeC. Our evaluation shows that the most significant
contribution to the  photon distribution in a nucleus can be
accurately evaluated  using  Eq. \ref{P}.
At the same time corrections for the incoherent effects maybe included as well, 
and may lead to  significant  corrections in the processes with nucleus break up.
In particular, this contribution   is of relevance for calculating the cross section of 
ultraperipheral processes where one often considers the cross section summed up over all final nuclear states or even just selects the final states where nucleus was excited \cite{Baltz:2007kq}.
  To a first approximation, to treat these contributions it is necessary to modify the standard Weizsacker - Williams formula for the photon flux by replacing the  the factor $Z^2F_A^2(t)$ by $ Z(Z-1)F_A^2(t) +ZF_N^2(t)$ (a more detailed treatment would involve including also transitions like $N\to \gamma +\Delta$). These issues will be considered  elsewhere.

\section*{Acknowledgements}
We thank  W.Vogelsang informing us  about   early studies of the effect of the Coulomb field 
in nucleons.  We also thank C.~Ciofi degli Atti and L.~Kaptari 
for discussions. The research was supported by DOE and BSF.

\end{document}